\documentclass[twocolumn,preprintnumbers,amsmath,amssymb,nofootinbib,showpacs]{revtex4-1}

\usepackage{mathrsfs}
 \usepackage{graphicx}
\usepackage{bm}
\usepackage{hyperref}
\usepackage[usenames]{color}
\usepackage{url}
\usepackage[normalem]{ulem}

\hypersetup{
    colorlinks=true,
    linkcolor=red,
    citecolor=red,
}

\newcommand{\remove}[1]{}

\def\be{\begin{equation}}
\def\ee{\end{equation}}
\def\ba{\begin{eqnarray}}
\def\ea{\end{eqnarray}}
\newcommand{\Z}[1]{\mathrm{Z}_{#1}}
\newcommand{\SU}[1]{\mathrm{SU}(#1)}
\newcommand{\U}[1]{\mathrm{U}(#1)}

\begin{document}

\title{Magnetic monopole - domain wall collisions}

\author{Micah Brush$^{1}$, Levon Pogosian$^{1}$, and Tanmay Vachaspati$^{2,3}$}

\affiliation{$^{1}$Department of Physics, Simon Fraser University, Burnaby, BC, V5A 1S6, Canada}

\affiliation{$^{2}$Department of Physics, Arizona State University, Tempe, Arizona 85287, USA}

\affiliation{$^{3}$Department of Physics, Washington University, St. Louis, MO 63130, USA}

\begin{abstract}
Interactions of different types of topological defects can play an important 
role in the aftermath of a phase transition.
We study interactions of fundamental magnetic monopoles and stable domain walls in a Grand Unified
theory in which $\SU5 \times \Z2$ 
symmetry is spontaneously broken to $\SU3 \times \SU2 \times \U1 / \Z6$.
We find that there are only two distinct outcomes depending on the relative orientation of the monopole and 
the wall in internal space. In one case, the monopole passes through the wall, while in the other it unwinds 
on hitting the wall. 

\end{abstract}


\maketitle

\section{Introduction}

Grand Unified Theories (GUTs) are based on large symmetry groups, the smallest of which is an $\SU5$ model
with an additional, possibly approximate, $\Z2$ symmetry.
When such large symmetries are broken in a cosmological setting, several kinds of topological defects 
can be produced. The ensuing cosmology will depend critically on the interactions of the different
defects. In particular, the $\SU5 \times \Z2$ symmetry breaking leads to the existence of magnetic 
monopoles and domain walls in the aftermath of the phase transition.
We expect the magnetic monopoles to interact with domain walls, potentially resolving the magnetic
monopole over-abundance problem \cite{Dvali:1997sa}. To investigate this idea further,
we study the interactions of $\SU5$ monopoles and $\Z2$ domain walls in this paper.

The interaction of monopoles and domain walls was also studied in \cite{Pogosian:1999zi} with the domain wall
structure given by
\begin{equation}
\Phi = \tanh \left ( \frac{z}{w} \right ) \Phi_0
\label{q=0wall}
\end{equation}
where the order parameter $\Phi$ is in the adjoint representation of $\SU5$, $\Phi_0$ is its constant 
vacuum expectation value (VEV), and
$w$ is the width of the domain wall. By numerical evaluation it was found that monopoles hitting
this domain wall will unwind and spread on the wall. Subsequently, however, it was found 
\cite{Pogosian:2000xv,Vachaspati:2001pw,Pogosian:2001fm,Vachaspati:2003zp}
that the model actually has several domain wall solutions, including the one in Eq.~(\ref{q=0wall}), 
and that the lightest (stable) wall has a different structure (see Sec.~\ref{sec:wall}). Hence the 
interaction of the stable wall and the monopole needs to be revisited.

In Sec.~\ref{model} we provide details of the $\SU5 \times \Z2$ model, the monopole solution,
the wall solutions, and finally our scheme for setting up a configuration with a monopole and
a domain wall together. This provides us with initial conditions that we numerically evolve in
Sec.~\ref{evolution}. The complexity of the field equations and the problem requires some special 
numerical techniques that we briefly 
describe in Sec.~\ref{evolution}. 

Our results are summarized in Sec.~\ref{conclusions}. Essentially we find that there are two
internal space polarizations for the monopole with respect to the wall. One of the polarizations
is able to pass through the wall with only some kinematic changes.
The monopole with the other polarization is unable to pass through the domain
wall and unwinds on the wall, radiating away its gauge fields. The disappearance of this
monopole is further explained in Sec.~\ref{conclusions}.

\section{The Model}
\label{model}

The $\SU5$ model we consider is given by the Lagrangian:
\begin{equation}
L = -{1\over 4} X^a_{\mu\nu}X^{a\mu\nu} +  {1\over 2} D_\mu\phi^a D^\mu\phi^a  - V(\Phi )
\label{lagrangian}
\end{equation}
where $\Phi = \phi^a T^a$ ($a = 1,...,24$),  $X^a_{\mu \nu}$ are the gauge field strengths defined as 
\begin{equation}
	X_{\mu \nu} = \partial_\mu X_\nu - \partial_\nu X_\mu - i g [X_\mu, X_\nu] \,,
\end{equation}
$X_\mu = X^a_\mu T^a$ are the gauge fields and $g$ is the coupling constant. $T^a$ are the generators 
of $\SU5$ normalized by Tr($T^aT^b$) = $\delta_{a b}/2$. The covariant derivative is given by
\begin{equation}
D_\mu \phi^a = \partial_\mu \phi^a - i g [X_\mu ,\Phi]^a  \,.
\label{covariantderivative}
\end{equation}
The most general renormalizable $\SU5$ potential is
\begin{equation}
V(\Phi ) = -m^2 {\rm Tr}\Phi^2 + \gamma {\rm Tr}\Phi^3
           +  h ({\rm Tr} \Phi^2 )^2 +
               \lambda {\rm Tr} \Phi^4 -V_0 \,,
\label{potential}
\end{equation}
and we will assume that $\gamma$ vanishes, giving the model an additional $\Z2$ 
symmetry. For $\lambda \ge 0$ and $h+7\lambda /30 \ge 0$, the potential has its global minimum at \cite{Ruegg:1980gf}
\begin{equation}
\label{eq.vev}
	\Phi_0 = \frac{\eta}{2\sqrt{15}} \text{diag}(2,2,2,-3,-3),
\end{equation}
with $\eta=m/\sqrt{h+7\lambda /30}$. The VEV, $\Phi_0$,
spontaneously breaks the $\SU5$ symmetry to $\SU3\times \SU2 \times \U1/ \Z6$.

In what follows, the four diagonal generators of $\SU5$ are chosen to be
\begin{equation}
\begin{split}
	\lambda_3 &= \frac{1}{2} \text{diag}(1,-1,0,0,0), \\
	\lambda_8 &= \frac{1}{2 \sqrt{3}} \text{diag}(1,1,-2,0,0), \\
	\tau_3 &= \frac{1}{2} \text{diag}(0,0,0,1,-1), \\
	Y &= \frac{1}{2\sqrt{15}} \text{diag}(2,2,2,-3,-3).
\end{split}
\end{equation}
We use $a=1,2,3$ to denote generators $T^a = \tau_a = \text{diag}(0,0,0,\sigma_a/2)$ where $\sigma_a$ are the Pauli spin matrices.

\subsection{The monopole}
\label{sec:monopole}

\begin{figure}[tbh]
\hspace*{-0.025\textwidth}\includegraphics[width=0.55\textwidth]{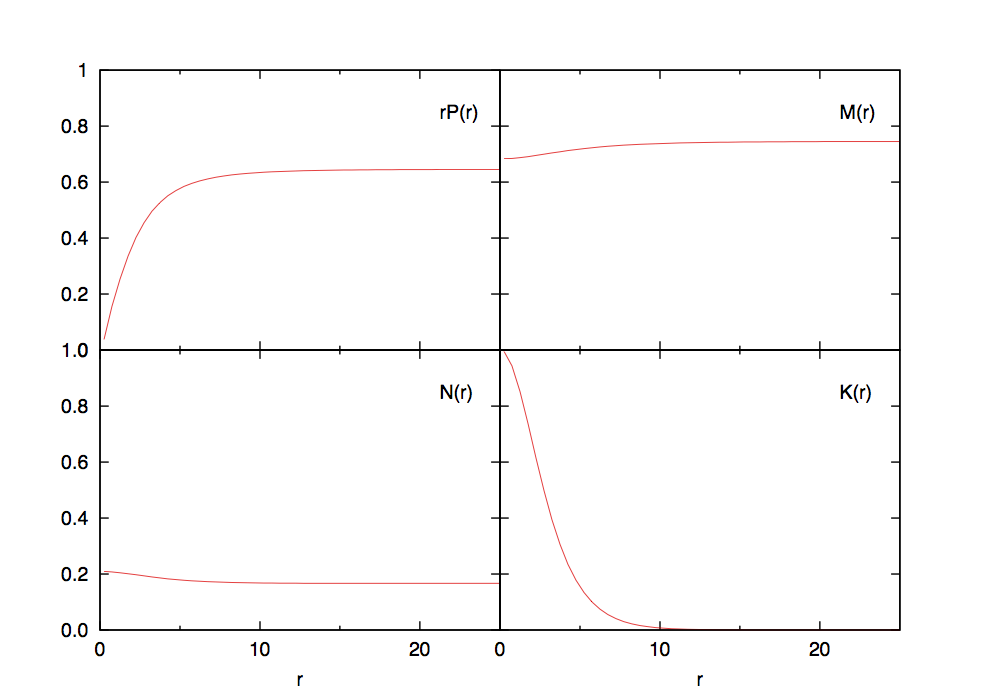}
\caption{The profile functions for the monopole alone, evaluated numerically, for a model with $\eta = 1$, 
$h/\lambda = -0.2$ and $\lambda=0.5$.}
\label{fig.monparam}
\end{figure}

Let us consider a magnetic monopole whose winding lies in the 4-5 block of $\Phi$. This is possible \cite{Wilkinson:1977yq} if we take the VEV along one of the radial directions far away from the monopole to be
\ba
\label{eqMinf}
\Phi_\infty &=& \frac{\eta}{2\sqrt{15}} \text{diag}(2,-3,2,2,-3) \nonumber \\ 
&=& \eta \sqrt{\frac{5}{12}}(\lambda_3 + \tau_3) + \frac{\eta}{6}(Y-\sqrt{5}\lambda_8).
\ea
The monopole ansatz for the scalar field can be written as \cite{Pogosian:2000xv}
\begin{equation}
\label{eq.phimonopole}
\Phi_M(r) = P(r) \sum_{a=1}^3 x^a \tau_a + M(r) \left( \frac{\sqrt{3}}{2}\lambda_3 - \frac{1}{2}\lambda_8 \right) + N(r) Y,
\end{equation}
while the non-zero gauge fields can be written as
\begin{equation}
	X_i^a = \epsilon_{ij}^a  \frac{x^j}{g r^2} (1-K(r)) , \ (a=1,2,3)
\label{eq.gaugemonopole}
\end{equation}
and $P(r), M(r), N(r),$ and $K(r)$ are profile functions that depend only on the spherical radial coordinate $r=\sqrt{x^2+y^2+z^2}$ and satisfy the boundary conditions:
\ba
\label{eq.profilebc}
\lim_{r \rightarrow \infty} r P(r)  &=& \eta \sqrt{\frac{5}{12}}, 
\ M(\infty) = \eta \frac{\sqrt{5}}{3}, \nonumber
\\ N(\infty) &=& \frac{\eta}{6},~~K(\infty) = 0.
\ea
The profile functions for the monopole alone were evaluated numerically and are shown in Fig.~\ref{fig.monparam}.

The non-Abelian magnetic field can be defined as \cite{Dokos:1979vu}
\[
B^k = -\frac{1}{2} \epsilon^{i j k} X_{ij}
\]
with the associated energy density given by ${\rm Tr} (B_k B^k)$. Far away from the centre, the monopole field becomes $B^k \rightarrow Q x^k /(g r^3)$, with $Q = \tau^j {x}^j /r$.

The monopole charge $Q$ includes a component along the generator of the unbroken $\U1$ symmetry 
($\Phi_\infty$ of Eq.~(\ref{eqMinf})), as well as $\SU2$ and $\SU3$ magnetic charges. The $\U1$ part of 
the magnetic field, which is a defining feature of a topological $\SU5$ monopole, is given by
\begin{equation}
B^k_Y = -\frac{1}{2} \epsilon^{i j k}  X^a_{ij} {\hat \phi}^a
\label{eq.bfield}
\end{equation}
where ${\hat \phi}^a \equiv \phi^a / \sqrt{\phi^b \phi^b}$. As discussed in \cite{'tHooft:1974qc},  other
definitions of the Abelian magnetic field are possible, and these differ from our definition but only within 
the core of the monopole. Since we only use our definition to plot the long range Abelian magnetic field
(see Fig.~\ref{fig.magnetic}) the definition in Eq.~(\ref{eq.bfield}) is sufficient.

\subsection{The wall}
\label{sec:wall}

Without loss of generality \cite{Vachaspati:2001pw}, the domain wall solution can be taken to be diagonal at all $z$ and written in terms of the diagonal generators of $\SU5$ as
\begin{equation}
\label{eq.dw}
	\Phi_{DW} (z) = a(z) \lambda_3 + b(z) \lambda_8 + c(z) \tau_3 + d(z) Y \,.
\end{equation}
In each of the two disconnected parts of the vacuum manifold ${\cal M}$ there are a total of 10 different diagonal VEVs corresponding to all possible permutations of 2's and 3's in Eq.~(\ref{eq.vev}). Topology dictates that there must be a domain wall separating any pair of VEVs from the two disconnected parts of ${\cal M}$. However, not every such pair of VEVs corresponds to a stable domain wall solution. For instance, as shown in \cite{Pogosian:2000xv}, the wall across which $\Phi_0$ goes to $-\Phi_0$ is unstable and will decay into a lower energy stable wall. The stable domain walls are obtained when both 3's in Eq.~(\ref{eq.vev}) change into 2's across the wall. 

\begin{figure}[tbh]
\hspace*{-0.025\textwidth}\includegraphics[width=0.55\textwidth]{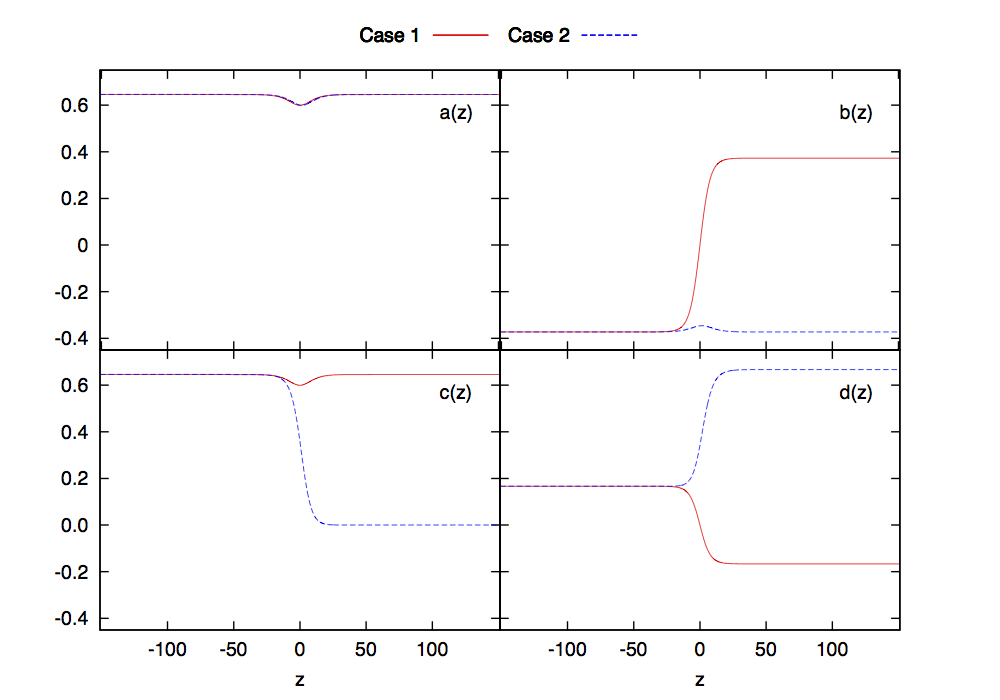}
\caption{The wall profile functions for Cases 1 and 2 for a model with $\eta = 1$, $h/\lambda = -0.2$ and $\lambda=0.5$. Note that the profile function $c(z)$ goes to zero in Case 2, which gives an unbroken $\SU2 \subset \SU3$ symmetry in the 4-5 block of $\Phi^{(2)}_+$. The profile function $a(z)$ is the same for Cases 1 and 2.}
\label{fig.wallparam}
\end{figure}

Let us choose the boundary condition at $z = -\infty$ to be
\ba
\label{eq.-inf}
	\Phi_- &=& \Phi(z = -\infty) = \frac{\eta}{2\sqrt{15}} \text{diag}(2,-3,2,2,-3) \nonumber \\ 
	&=& \eta \sqrt{\frac{5}{12}}(\lambda_3 + \tau_3) + \frac{\eta}{6}(Y-\sqrt{5}\lambda_8) \,.
\ea
For this choice of $\Phi_-$, there are three different choices of $\Phi(z=+\infty)$, proportional to 
\ba
\nonumber
&&\text{diag}(3,-2,-2,3,-2) \\
\nonumber
&&\text{diag}(-2,-2,3,3,-2) \\
&&\text{diag}(3,-2,3,-2,-2) \,,
\label{3VEVs}
\ea
that lead to stable domain walls. For the purpose of understanding the monopole-wall interactions, it is sufficient to consider only two of the above, corresponding to the two 
distinct entries in the 4-5 block of $\Phi$. We take 
the first to be the same as in \cite{Pogosian:2000xv}, subsequently referred to as Case 1:
\ba
\label{eq.inf1}
\Phi^{(1)}_+ &=& \frac{\eta}{2\sqrt{15}} \text{diag}(3,-2,-2,3,-2) \nonumber \\ 
&=& \eta \sqrt{\frac{5}{12}}(\lambda_3 + \tau_3) - \frac{\eta}{6}(Y-\sqrt{5}\lambda_8) \,.
\ea
The value of the field in the core of this wall is proportional to $\text{diag}(1,-1,0,1,-1)$. The other case, subsequently referred to as Case 2, has
\ba
\label{eq.inf2}
\Phi^{(2)}_+ &=& \frac{\eta}{2\sqrt{15}} \text{diag}(3,-2,3,-2,-2) \nonumber \\
&=& \eta \frac{\sqrt{15}}{6}\lambda_3 + \frac{\eta}{6}(4Y-\sqrt{5}\lambda_8) \,,
\ea
with the field in the wall being proportional to $\text{diag}(1,-1,1,0,-1)$. 
A novel feature
of these walls is that the unbroken symmetry groups on either side of the wall are isomorphic
to each other but they are realized along different directions of the initial $\SU5$ symmetry
group. Hence the wall is the location of a clash of symmetries \cite{Davidson:2002eu}.

Note that the symmetry within the wall is [$\SU2 \times \U1$]$^2$.  
The $\SU2$'s correspond to rotations in the 1-3 and 2-5 blocks and the $\U1$'s
to rotations along $\sigma_3$ in the 1-2 and 3-5 blocks. Therefore the symmetry group within
the wall is 8-dimensional, and is smaller than the 12-dimensional symmetry outside the 
wall\footnote{For simplest domain walls, such as kinks in $\lambda \Phi^4$, the full 
symmetry of the Lagrangian is restored inside the core. However, the symmetry inside stable domain walls in 
$\SU{N} \times \Z2$ is always lower than that of the vacuum \cite{Vachaspati:2001pw}.}.
Also note that the symmetry in the 4-5 block is different for the $\Phi^{(1)}_+$ and $\Phi^{(2)}_+$ vacua. 
This is going to be of direct relevance for the fate of the monopoles.

The profile functions $a(z)$, $b(z)$, $c(z)$ and $d(z)$ for both cases are shown in Fig.~\ref{fig.wallparam}. In each case, they are linear combinations of two functions $F_+(z)$ and $F_-(z)$ defined by the alternative way of writing the domain wall solution \cite{Pogosian:2000xv}
\begin{equation}
\Phi_{DW} = \frac{\Phi_+(z) - \Phi_-(z)}{2} F_-(z) + \frac{\Phi_+(z) + \Phi_-(z)}{2} F_+(z) \, ,
\end{equation}
where $F_+(\pm \infty)=1$, $F_-(\pm \infty) = \pm 1$. For a general choice of parameters, functions $F_\pm(z)$ must be found numerically. For $h/\lambda = -3/20$, they are known in closed form \cite{Pogosian:2000xv}: $F_+ (z) = 1$, $F_-(z) = \tanh (mz/\sqrt{2})$. Correspondingly, for this value of $h/\lambda$, the four functions $a(z)$, $b(z)$, $c(z)$, and $d(z)$ are either constant or describe a transition from one constant value to another. For $h/\lambda \ne -3/20$ the ``constant'' functions develop a small bump around $z=0$ as can be seen in Fig.~\ref{fig.wallparam}.

\subsection{Monopole and Wall}

As our initial configuration, we take the monopole to be on the $z=-\infty$ side, far away from the wall. In this case, the ansatz for the initial combined field configuration of the wall and the monopole can be written as \cite{Pogosian:2000xv}
\ba
\label{eq.phimdm}
\Phi_{M+DW} &=& P(r) \frac{c(z')}{c(-\infty)} \sum_{a=1}^3 x^a \tau_a + N(r) \frac{d(z')}{d(-\infty)} Y \nonumber \\
&+& M(r) \left( \frac{\sqrt{3}}{2} \frac{a(z')}{a(-\infty)} \lambda_3 - \frac{1}{2} \frac{b(z')}{b(-\infty)} \lambda_8 \right)
\ea
where $z' = \gamma (z- z_0)$, $\gamma=1/\sqrt{1-v^2}$ is the boost factor, $v$ is the wall velocity and $z_0$ is the initial position of the wall. The monopole is at $x=0=y=z$.   It is easy to check that, far away from the monopole, the profile functions take on the values in Eq.~(\ref{eq.profilebc}) and $\Phi_{M+DW} \rightarrow \Phi_{DW}$. Close to the monopole, $z' \rightarrow -\infty$, since the monopole is initially very far from the wall, and $\Phi_{M+DW} \rightarrow \Phi_M$ as desired. We work in the temporal gauge, $X_0^a=0$, and with the initial ansatz for the gauge fields given by Eq.~(\ref{eq.gaugemonopole}) for both cases.

It is instructive to examine the difference in the nature of the magnetic field in Cases 1 and 2. As mentioned in Sec.~\ref{sec:monopole}, the charge of our monopole along the $z$-direction, $Q = (1/2)\text{diag}(0,0,0,1,-1)$, is a combination of the $\U1$, the $\SU2$, and the $\SU3$ magnetic charges. Since the VEV of $\Phi$ in our model is along the generator of (hypercharge) $\U1$, the magnetic field, 
as defined in Eq.~(\ref{eq.bfield}), corresponds solely to the $\U1$ component of the charge.
In Case 1, ${\rm Tr}(Q \Phi)$ is the same on both sides of the wall and the $\U1$ magnetic field is unaffected by the presence of the DW. In Case 2, however, ${\rm Tr}(Q \Phi^{(2)}_+)=0$ and there is no magnetic field corresponding to the unbroken $\U1$ on the $z=+\infty$ side of the wall. Instead, the gauge field on that side is associated with an 
$\SU2$ subgroup of the unbroken $\SU3$. We note that, while the magnetic \emph{energy density} associated with 
the gauge field is unaffected by the presence of the wall, it is specifically the $\U1$ magnetic field that is a defining feature of a topologically stable monopole.

\begin{figure}[tbh]
\hspace*{-0.025\textwidth}\includegraphics[width=0.55\textwidth]{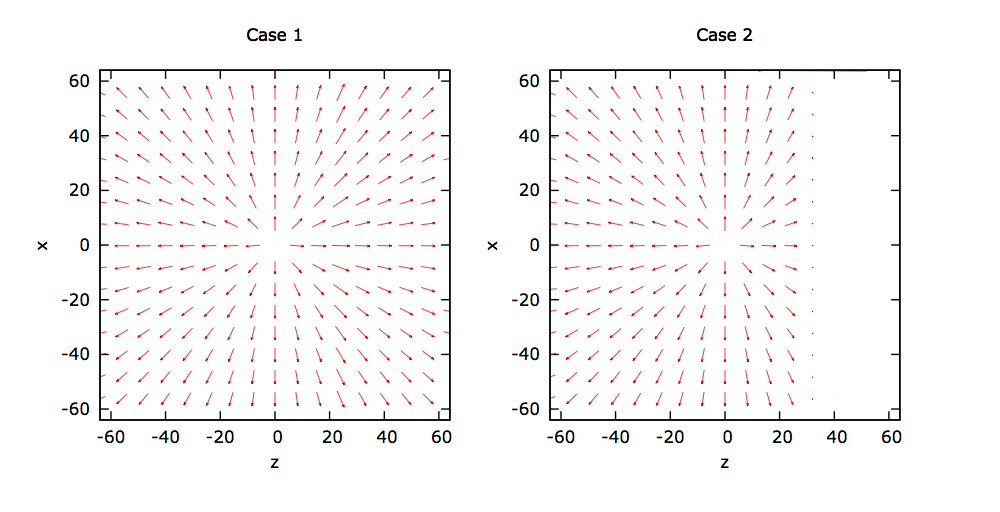}
\caption{The magnetic field ${\bf B}_Y$ (defined in Eq.~(\ref{eq.bfield})) multiplied by $r^2$ for Cases 1 and 2,  where at each point $r^2B_Y^z$ and $r^2B_Y^x$ are plotted as a vector. In Case 1, there is a magnetic field associated with the unbroken $\U1$ symmetry on both sides of the wall. In Case 2, the magnetic field becomes associated with the $\SU2 \subset \SU3$ on the $z=+\infty$ side on the wall, while its $\U1$ component vanishes. Note that it is the $\U1$ magnetic field that characterizes a topologically stable monopole.
}
\label{fig.magnetic}
\end{figure}

The magnetic field, as defined in Eq.~(\ref{eq.bfield}), is plotted for both cases in Fig.~\ref{fig.magnetic}, where the vectors have components $r^2 B_Y^z$ and $r^2 B_Y^x$. This plot shows that, in Case 1, there is a $\U1$ magnetic field on both sides of the wall falling off as $r^2$ as expected, while in Case 2 the $\U1$ magnetic field is zero on the $z=+\infty$ side of the wall.

\section{Evolution}
\label{evolution}

\begin{figure*}[htb]
\includegraphics[width=0.7\textwidth]{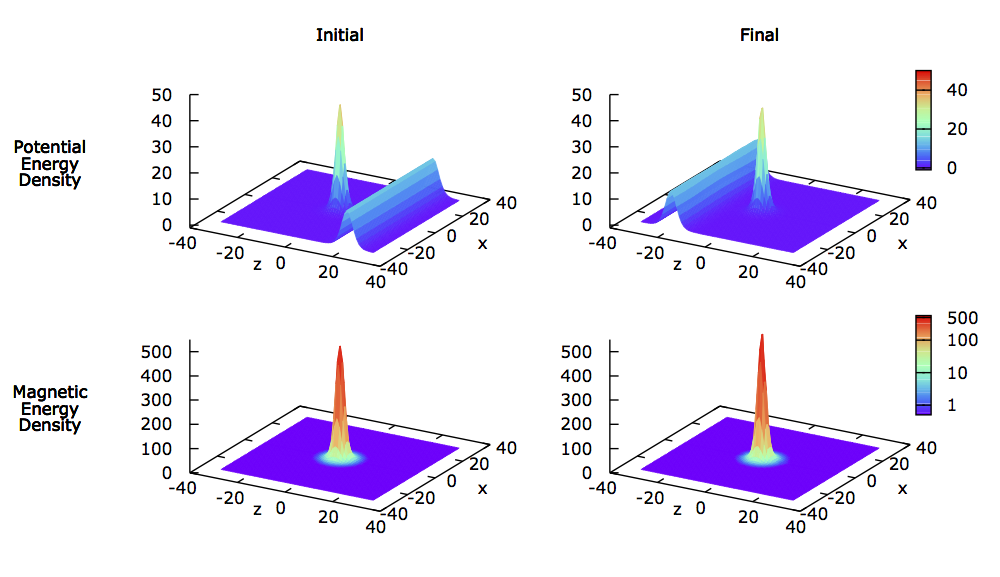}
\caption{The potential and magnetic energy densities in the $xz$ plane for the colliding monopole and wall in Case 1. We see that the monopole passes through the wall and the energy densities remain localized. Additionally, we see the magnetic energy density is unchanged before and after the collision.}
\label{fig.case1mp}
\end{figure*}

\begin{figure}[tbh]
\hspace*{-0.025\textwidth}\includegraphics[width=0.55\textwidth]{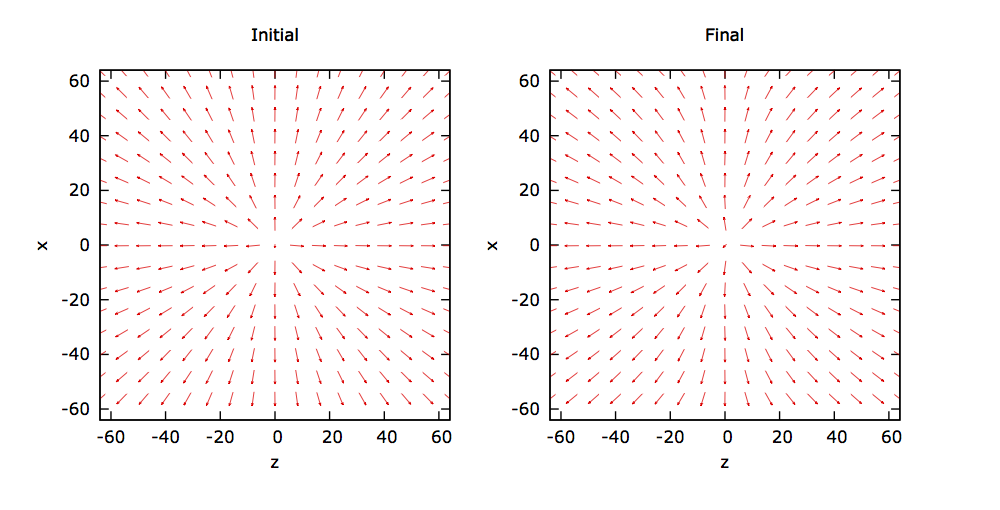}
\caption{The scalar field $\phi^a$ in the $xz$ plane for the colliding monopole and wall in Case 1, where at each point $\phi^3$ and $\phi^1$ are plotted as a vector. In this case the scalar field arrangement in direction and magnitude remains virtually unchanged.}
\label{fig.case1phi}
\end{figure}

Let us consider an initial monopole-wall configuration given by Eq.~(\ref{eq.phimdm}) in which VEV at $z=-\infty$ is given by $\Phi_-$ in Eq.~(\ref{eq.-inf}). As mentioned in the previous Section, there are 2 types of boundary conditions at $z=+\infty$, given by Eqs.~(\ref{eq.inf1}) and (\ref{eq.inf2}), dubbed Case 1 and Case 2, leading to 2 different outcomes of the monopole-wall collision.

Before considering the two cases in detail, let us note that initially, when the monopole and the wall are 
very far away from each other, the field configuration has just three non-zero gauge fields and six scalar 
fields corresponding to the generators that appear in Eq.~(\ref{eq.phimdm}). Because these six generators 
form a closed algebra, it follows from the equations of motion that the subsequent evolution does 
not involve fields corresponding to the other 18 generators. Namely, the scalar and the gauge field equations are
\ba
D_\mu D^\mu \phi^a &=& - {\partial V / \partial \phi^a} 
\label{eom:scalar}
\\
D_\mu X^{\mu \nu a} &=& g f_{a b c} (D^\nu \Phi)^b \phi^c
\label{eom:gauge}
\ea
where $f_{a b c}$ are the $\SU5$ structure constants defined by $[T^a,T^b] = i f_{a b c} T^c$. Let $\cal{C}$ be the set of indices of the 6 generators that appear in the initial field configuration given by Eq.~(\ref{eq.phimdm}). Since the 6 generators form a closed algebra,
$f_{abc} = 0$ for $a \notin \cal{C}$ and $b,c \in \cal{C}$. Now let $\phi^a$ and $X_\mu^a$ be fields corresponding to any $a \notin \cal{C}$. If $\phi^a$ and $X_\mu^a$ are zero at the initial time, they will remain zero if $f_{abc} = 0$ for $b,c \in \cal{C}$ and ${\partial V / \partial \phi^a} \ne 0$. The former condition is satisfied as mentioned above, while the latter holds since ${\rm Tr}[T^aT^b] \propto \delta_{ab}$ and ${\rm Tr}[T^aT^bT^cT^d] = 0$ for $b,c,d \in \cal{C}$, as we have checked by explicit evaluation. Thus, for our purposes, it is sufficient\footnote{Although the field components for $a \not\in \cal{C}$ continue to vanish during evolution if they vanish initially, we cannot exclude the possibility that the fields in these other directions may grow unstably if they did not vanish initially.} to consider only $a \in \cal{C}$.

Our numerical implementation is based on techniques developed in \cite{Pogosian:1999zi}. First, the DW and the monopole profile functions are found via numerical relaxation. The monopole is initially located at the center of
the lattice. We give the DW a velocity towards the monopole and boosted profiles are inserted into the initial configuration given by Eq.~(\ref{eq.phimdm}). With the initial time derivatives simply determined from the Lorentz boost factor, this initial configuration is evolved forward in time using a staggered leapfrog code. The boundary conditions require special care since the wall extends all the way across the lattice. We have implemented boundary conditions in which the field is extrapolated across the boundary. We have
numerically tested that this boundary condition leads to a smoothly evolving domain wall, without any
spurious incoming radiation. Even though our problem has axial symmetry, we work in Cartesian coordinates as this offers superior stability.
 However, as discussed in \cite{Alcubierre:1999ab}, we take advantage of the axial symmetry of our configuration 
 to restrict the lattice to just three lattice spacings along the $y$ direction. We then use a $256 \times 256$ lattice grid for the $x$ and $z$ coordinates. Additionally, the axial symmetry allows us to solve only for positive $x$ and use reflection to find the fields at negative $x$. The units of length are set by $\eta=1$ and we take each lattice spacing to correspond to half of a length unit. In these units, the range of $x$ and $z$ axis for a $256 \times 256$ grid is $[-64,64]$. Note that in some figures we do not plot the entire lattice. The radius of the monopole core is about $10$ length units and is about the same as a half of the domain wall width. At the initial time, the wall is 30 length units away from the center of the monopole.
	
\subsection{Case 1: the monopole passes through}

It is not difficult to predict that the monopole in Case 1 will pass through the wall. The monopole winding is due to the fields in the $\SU2$ subgroup corresponding to generators $\tau_a$, $a=1,..,3$. In Eq.~(\ref{eq.phimdm}), these fields are multiplied by the function $c(z)$ which has the same value at $z=\pm \infty$ and, as known from \cite{Pogosian:2000xv}, is approximately constant across the domain wall. Only $b(z)$ and $d(z)$ change signs across the wall, but these are irrelevant for the winding of the monopole. Thus, the presence of the wall is of no qualitative consequence to the winding of the monopole or its profile functions. The only effect is the small change in $c(z)$ around $z=0$ (note that, as mentioned earlier, $c(z)$ is strictly a constant when $h/\lambda = -3/20$).

We numerically collide the monopole and the wall by giving the wall an initial velocity of $0.8$ (in speed of light units) and choosing parameters $\eta = 1$, $h = -\lambda/5$, and $\lambda=0.5$ for $V(\Phi)$.

Fig.~\ref{fig.case1mp} shows the potential and magnetic energy densities as the wall hits the monopole in Case 1. In addition, we plot the scalar field configuration in Fig.~\ref{fig.case1phi}, where each point is a vector with components $\phi^3$ and $\phi^1$. These figures show that the magnetic energy density and the scalar field configuration remain unchanged after the collision, and that the potential energy densities corresponding to the monopole and the domain wall remain localized. This does not imply a complete absence of interaction between the wall and the monopole -- some interaction is expected due the non-linearity of the scalar field potential.

To see if the monopole gains momentum due to the interaction, we have evaluated the centre of energy (COE) defined as
\be
z_{\rm COE}(t) = { \int_V d^3x \ z \ \rho(t,{\bf x}) \over \int_V d^3x \ \rho(t,{\bf x})} \ ,
\ee
where $V$ is the volume of a finite cylindrical region centred at the origin and extending $1/8$th of the lattice size in the $x$- and $z$-directions, while $\rho$ is the energy density. For $h/\lambda=0$, we give the wall a velocity of $v=0.9$ towards the monopole and compare the initial $z_{\rm CEO}$ to the one after the wall passes away. We see a very slow drift of the COE in the direction of the wall velocity. We performed the same procedure using different model parameters and wall velocities and the outcome was qualitatively the same. In all cases, while the direction of the drift is clear, the magnitude is extremely small and too close to the numerical uncertainties to allow a definitive quantitative analysis.

\subsection{Case 2: the monopole unwinds}

\begin{figure*}[tbh]
\includegraphics[width=\textwidth]{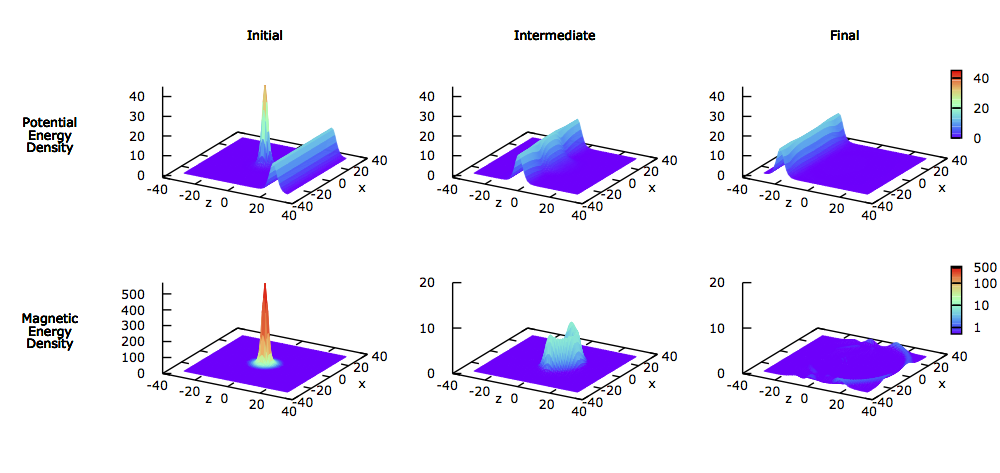}
\caption{The potential and magnetic energy densities in the $xz$ plane for the colliding monopole and wall in Case 2. We can see that as the domain wall and monopole collide, the potential energy contained by the monopole disappears and the monopole begins to radiate away its magnetic energy in a hemispherical wave. Note that the middle and final plots for the magnetic energy density have a much smaller scale as the ripples are not visible at the original scale.}
\label{fig.case2mp}
\end{figure*}

\begin{figure*}[tbh]
\includegraphics[width=\textwidth]{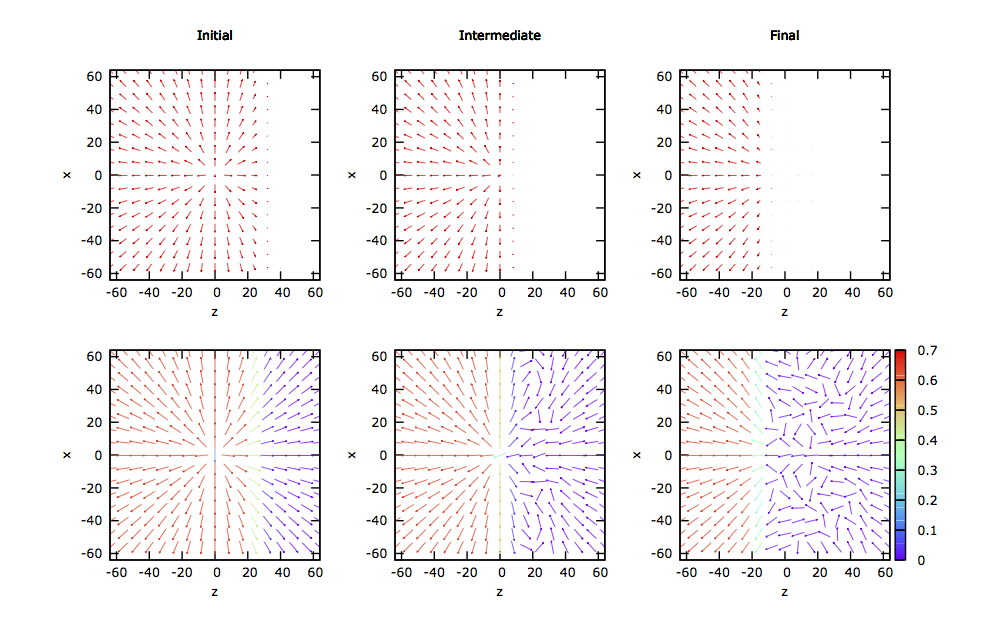}
\caption{The scalar field $\phi^a$ in the $xz$ plane for the colliding monopole and wall in Case 2. At each point in the first row, $\phi^3$ and $\phi^1$ are plotted as a vector. In the second row, the length of the arrow is fixed, while the direction of the arrow is given by $\tan^{-1}(\phi^3/\phi^1)$ and the color represents the magnitude of the field $|\phi| = \sqrt{\phi^a\phi^a}$ for $a=1,2,3$. The first row shows the monopole unwinding as the wall sweeps past it, and the second shows how the fields arrange themselves to unwind the monopole.
}
\label{fig.case2phi}
\end{figure*}

As in Case 1, it is possible to guess the outcome of the monopole-wall collision without doing numerical simulations. For this, we note that $\Phi^{(2)}_+$ has an $\SU2$ symmetry in the 4-5 block, which means that there is no topology that can support the winding. Thus, the monopole cannot exist in that corner of the matrix. An equivalent way to see this is to note that the function $c(z)$, which multiplies the three relevant monopole scalar fields, goes to zero at $z=+\infty$ (see Fig.~\ref{fig.wallparam}), effectively erasing the monopole.

Additional insight can be gained by noting that the long range
magnetic field of the monopole transforms into an $\SU3$ magnetic field on the far side of
the wall. More explicitly, the $\U1$ magnetic field is given by Eq.~(\ref{eq.bfield}) with $X_{ij}^a$ 
determined using the solution in Eq.~(\ref{eq.gaugemonopole}). Since $X_{ij}$ only has components 
in the $\tau^a$ directions, it lies in the 4-5 block. However, the 4-5 block is entirely within the unbroken $\SU3$ 
on the right-hand side of the wall. Thus the long range magnetic field of the monopole is 
purely $\SU3$ on the right-hand side of the wall and, from the
vantage point of someone there, there is no $\U1$ magnetic field emerging
from the left-hand side of the wall. However, a $\U1$ magnetic field is an essential feature of a topological 
monopole. Thus, from the right-hand side of the wall, there is no magnetic monopole in the system,
only some source of $\SU3$ magnetic flux. 

Doing the numerical simulation with the parameters chosen as before, we plot the potential and magnetic energy densities as the wall hits the monopole in Fig.~\ref{fig.case2mp}. This figure shows that the potential energy for the monopole disappears as the wall and monopole collide, and the magnetic energy that was stored in the monopole radiates away in a hemispherical wave. The collision was simulated with initial wall velocities ranging from $0.1$ to $0.99$ for $h/\lambda = -1/5$, and initial wall velocities of $0.6$, $0.8$ and $0.99$ for $h/\lambda = -3/20$ and $1/5$. In all of these cases, the result of the collision was unchanged.

In Fig.~\ref{fig.case2phi}, we show the $a=1,2,3$ components of the scalar field using two different representations. In the first row, the fields $\phi^3$ and $\phi^1$ are plotted as a vector. The plot shows that the components of the field that are responsible for the winding vanish on the $z=+\infty$ side of the wall. In the second row of Fig.~\ref{fig.case2phi}, the color represents the magnitude $|\phi| \equiv \sqrt{\phi^a\phi^a}$, $a=1,2,3$, while vectors are drawn of fixed length and direction  given by $\tan^{-1}(\phi^3/\phi^1)$. Even though $|\phi|$ becomes very small, it is not strictly zero at a finite distance from the wall, and so one can still define the direction of the arrow in this way. One can see that initially the field has a hedgehog configuration across the wall. However, as the wall sweeps along, the fields on the $z=+\infty$ side of the wall rotate around in such a way as to unwind the monopole. In the final step, all fields that are non--zero are pointing in one direction, and therefore the monopole winding is gone.

\section{Conclusions}
\label{conclusions}

In a Grand Unified model there can be several types of defects, including magnetic monopoles and 
domain walls. In the aftermath of the cosmological phase transition in which the Grand 
Unified symmetry is spontaneously broken to the standard model symmetry, the monopoles and
walls will interact\footnote{Scattering of fermions and GUT domain walls was studied in \cite{Steer:2006ik}}. We have studied these interactions explicitly in an $\SU5 \times \Z2$
GUT, taking into account that the model has several different types of domain walls, and that only the lowest
energy wall is expected to be cosmologically relevant. Even this stable
wall has several different orientations in internal space, two of which are distinct for the purposes of
monopole-wall interaction. 

The first wall (Case 1 above) is found to be transparent to the monopole. This is simply because
the domain wall mainly resides in a certain block of field space, while the winding of the monopole 
resides in a different non-overlapping block. The interactions between the monopole and the wall are very weak, and only affect the dynamics of the monopole as it passes through the wall. Depending on the parameters, the monopole might be 
attracted or repelled by the wall leading to a time delay or advance as the monopole goes through.

The second wall (Case 2 above) is opaque to the monopole. When the monopole hits the
wall its energy is transformed into radiation on the other side of the wall, as seen in
Fig.~\ref{fig.case2mp}. A useful way to picture this system is to consider 
a magnetic monopole that is located inside a spherical domain wall.
Now there is a topological magnetic monopole inside the wall, but only an $\SU3$ magnetic 
flux from the outside. In particular, there is no topological magnetic monopole as seen from
the outside. Therefore the spherical wall itself must carry the topological charge of an 
antimonopole\footnote{The correspondence between spherical domain walls and global monopoles in $\SU{N}$ has previously been noted in \cite{Pogosian:2001fm}.}.
If the spherical wall shrinks, either it can annihilate the magnetic monopole within it and radiate away 
the energy, or the monopole can escape the wall, in which case the wall would then
collapse into an antimonopole so that the total topological charge of the system continues
to vanish. Our explicit numerical evolution shows that annihilation occurs for the parameter ranges 
we have considered. We note that the unwinding of the monopole in the Case 2 may be related to the mechanism of formation of non-Abelian clouds (massless monopoles) \cite{Lee:1996vz}.

Our results have bearing on cosmology as they explicitly show the possible destruction
of magnetic monopoles. 
In the case where the $\Z2$ symmetry is approximate, the walls will eventually
decay away, and it is possible that these interactions could lead to a universe that is free of magnetic monopoles.
Estimates in \cite{Dvali:1997sa} indicate that this possibility is worth investigating in more detail. 
With several types of domain walls and monopoles simultaneously forming in a phase 
transition \cite{Pogosian:2002ua,Antunes:2003be,Antunes:2004ir}, and with the complex nature of both the 
inter-wall \cite{Pogosian:2001pq} and monopole-wall interaction, the fate of the monopoles will remain uncertain 
until a comprehensive simulation of the GUT phase transition is performed. We leave this for a future study.

\acknowledgments

LP and MB are supported by the National Sciences and Engineering Research Council of Canada.
TV gratefully acknowledges the Clark Way Harrison Professorship at Washington University during the
course of this work, and was supported by the DOE at ASU.


\begin{thebibliography}{9}

\bibitem{Dvali:1997sa} 
  G.~R.~Dvali, H.~Liu and T.~Vachaspati,
  Phys.\ Rev.\ Lett.\  {\bf 80}, 2281 (1998)
  [hep-ph/9710301].

\bibitem{Pogosian:1999zi} 
  L.~Pogosian and T.~Vachaspati,
  Phys.\ Rev.\ D {\bf 62}, 105005 (2000)
  [hep-ph/9909543].

\bibitem{Pogosian:2000xv} 
  L.~Pogosian and T.~Vachaspati,
  Phys.\ Rev.\ D {\bf 62}, 123506 (2000)
  [hep-ph/0007045].

\bibitem{Vachaspati:2001pw} 
  T.~Vachaspati,
  Phys.\ Rev.\ D {\bf 63}, 105010 (2001)
  [hep-th/0102047].

\bibitem{Pogosian:2001fm} 
  L.~Pogosian and T.~Vachaspati,
  Phys.\ Rev.\ D {\bf 64}, 105023 (2001)
  [hep-th/0105128].

\bibitem{Vachaspati:2003zp} 
  T.~Vachaspati,
  Phys.\ Rev.\ D {\bf 67}, 125002 (2003)
  [hep-th/0303137].

\bibitem{Ruegg:1980gf} 
  H.~Ruegg,
  Phys.\ Rev.\ D {\bf 22}, 2040 (1980).

\bibitem{Wilkinson:1977yq} 
  D.~Wilkinson and A.~S.~Goldhaber,
  Phys.\ Rev.\ D {\bf 16}, 1221 (1977).
  
\bibitem{Dokos:1979vu} 
  C.~P.~Dokos and T.~N.~Tomaras,
  Phys.\ Rev.\ D {\bf 21}, 2940 (1980).
  
\bibitem{'tHooft:1974qc} 
  G.~'t Hooft,
  Nucl.\ Phys.\ B {\bf 79}, 276 (1974).

\bibitem{Liu:1996ea} 
  H.~Liu and T.~Vachaspati,
  Phys.\ Rev.\ D {\bf 56}, 1300 (1997)
  [hep-th/9604138].

\bibitem{Davidson:2002eu} 
  A.~Davidson, B.~F.~Toner, R.~R.~Volkas and K.~C.~Wali,
  Phys.\ Rev.\ D {\bf 65}, 125013 (2002)
  [hep-th/0202042].

\bibitem{Alcubierre:1999ab} 
  M.~Alcubierre, S.~Brandt, B.~Bruegmann, D.~Holz, E.~Seidel, R.~Takahashi and J.~Thornburg,
  Int.\ J.\ Mod.\ Phys.\ D {\bf 10}, 273 (2001)
  [gr-qc/9908012].

\bibitem{Steer:2006ik} 
  D.~A.~Steer and T.~Vachaspati,
  Phys.\ Rev.\ D {\bf 73}, 105021 (2006)
  [hep-th/0602130].

\bibitem{Lee:1996vz} 
  K.~M.~Lee, E.~J.~Weinberg and P.~Yi,
  Phys.\ Rev.\ D {\bf 54}, 6351 (1996)
  [hep-th/9605229].

\bibitem{Pogosian:2002ua} 
  L.~Pogosian and T.~Vachaspati,
  Phys.\ Rev.\ D {\bf 67}, 065012 (2003)
  [hep-th/0210232].

\bibitem{Antunes:2003be} 
  N.~D.~Antunes, L.~Pogosian and T.~Vachaspati,
  Phys.\ Rev.\ D {\bf 69}, 043513 (2004)
  [hep-ph/0307349].

\bibitem{Antunes:2004ir} 
  N.~D.~Antunes and T.~Vachaspati,
  Phys.\ Rev.\ D {\bf 70}, 063516 (2004)
  [hep-ph/0404227].

\bibitem{Pogosian:2001pq} 
  L.~Pogosian,
  Phys.\ Rev.\ D {\bf 65}, 065023 (2002)
  [hep-th/0111206].

\end{thebibliography}
\end{document}